\pdfoutput=1
\documentclass[sigconf]{acmart}

\copyrightyear{2026}
\acmYear{2026}
\setcopyright{cc}
\setcctype{by}
\acmConference[RecSysChallenge '26]{RecSys Challenge 2026}{October 02, 2026}{Minneapolis, MN, USA}
\acmBooktitle{RecSys Challenge 2026 (RecSysChallenge '26), October 02, 2026, Minneapolis, MN, USA}
\acmDOI{10.1145/3842413.3842427}
\acmISBN{979-8-4007-2863-1/2026/10}

\newcommand{\teamname}{sanjeevsuresh}
\newcommand{\codeurl}{\url{https://github.com/Sanjeev-S/recsys2026-request-audit}}
\newcommand{\scorenumbers}{composite 0.528, of which nDCG@20 0.440}

\begin{document}

\title[When the Label Ignores the Request]{When the Label Ignores the Request: Auditing Policy-Selected Targets in Synthetic Conversational Music Recommendation}

\author{Sanjeev Suresh}
\affiliation{%
  \institution{Independent Researcher}
  \city{San Francisco}
  \country{USA}
}
\email{sanjeev@sanjeevsuresh.com}

\begin{abstract}
Synthetic dialogues generated by LLM pipelines now serve as complete conversational-recommendation benchmarks: an LLM listener talks to an LLM recommender, and the track logged next in the conversation becomes the official label for each turn. These policy-selected labels make large-scale evaluation reproducible, but they are proxies for what the simulated user asked. We audit the one place where label and request are directly comparable: turns where the user asks for an exact song by name. In the RecSys Challenge 2026 TalkPlay benchmark, using visible dialogue and catalog metadata alone, we find that the official label contradicts the user's exact-song request in half of the audited development turns. This matters beyond one benchmark: naming the desired item is the dominant intent in real music search, where deployed systems avoid substituting an alternative for an exactly named item, on the premise that it costs satisfaction. A small training-time supplement closes most of the gap: adding catalog-resolved request-satisfying targets to a small fraction of training turns yields a 53.3\% relative gain in nDCG@20 on the 43 conflict turns while leaving the official metric intact, verified against a matched control that detects the same requests but trains only on official labels. Team: \teamname.
\end{abstract}

\ccsdesc[500]{Information systems~Recommender systems}
\ccsdesc[300]{Computing methodologies~Natural language processing}
\keywords{conversational recommendation, synthetic benchmarks, label validity, music recommendation, behavioral evaluation, RecSys Challenge}

\maketitle

\section{Introduction}
\label{sec:intro}

The RecSys Challenge 2026 asks systems to continue a conversational music-recommendation session: given the dialogue so far, rank tracks from a fixed catalog and write the next response~\cite{doh2026recsyschallenge}. The benchmark is synthetic, built by the TalkPlayData 2 pipeline to remedy the scarcity of large-scale conversational music data~\cite{choi2025talkplaydata2,doh2025talkplay}. One LLM plays a listener, another plays the recommender, and the track logged after each user turn becomes that turn's official label, which challenge systems must recover by ranking the full catalog, scored as nDCG@20. The label therefore records \emph{what the generation policy played next}, not necessarily \emph{what the user asked for} (Section~\ref{sec:benchmark} details the pipeline). In most turns there is no way to tell the two apart: a request like ``something moody for late-night driving'' admits many valid continuations, and the policy's choice is as defensible as any. We focus on the one family of turns where label and request are directly comparable: the user names an exact song.

In these turns the dialogue itself identifies a verifiable target. When a user asks ``Could you play the title track, `Mothership Connection (Star Child)'?'', catalog metadata resolves the request to exactly that track by Parliament. The official label for that turn, however, is \emph{Intimate Connection} by Kleeer. On the organizer development set we find this is not an outlier: our audit flags 82 turns with a visible exact-song directive, and in 41 of them the official label points to a different track. A system trained to predict official labels is therefore trained --- in half of the audited turns --- to override what the user literally asked for.

We argue this particular divergence is worth measuring and fixing, for a reason that extends beyond any single benchmark. Naming the desired item dominates real music interaction: most natural-language music queries are known-item searches~\cite{lee2010knownitem}; 81\% carry bibliographic metadata such as a title or performer~\cite{bainbridge2003music}; music playback is the most common command category on Amazon Alexa and among the top uses across voice assistants~\cite{ammari2019voice}; and users in a ``focused'' mindset judge a system by whether it returns the one item they have in mind~\cite{hosey2019justgive}. In production-scale experiments, a music streaming platform surfaces alternative content only on \emph{non-focused} queries, on the premise that substitution on focused queries costs satisfaction~\cite{tomasi2020query}. A conversational recommender that learns to ignore explicit song requests inherits a failure mode known to erode voice-assistant trust and use~\cite{luger2016pa}.

At the same time, not every visible request should become a training target. ``Play something from the 80s'' looks like a constraint, but release-date metadata is unreliable and the target set is not certifiable; rewriting labels from such patterns would replace one proxy with a worse one. We therefore act only where the evidence clears a strict bar, and report the five request families we screened and declined (Section~\ref{sec:families}). Our contributions are:

\begin{itemize}
  \item \textbf{A quantified label--request conflict audit}: on visible exact-song requests in the organizer development set, the official label conflicts with the request in 41/82 turns.
  \item \textbf{A guarded training supplement}: supplementing 1.5\% of training turns with request-satisfying targets lifts nDCG@20 on the exact/version conflict slice --- the turns where official label and resolved request diverge (Section~\ref{sec:training}) --- from 0.523 to 0.802 on the development set, while preserving the official metric, verified against a matched control (Table~\ref{tab:main}).
  \item \textbf{A released audit artifact}: ID-keyed annotations, the detector, the target-remap script that rebuilds the request-satisfying evaluation slice, and the model training scripts, so others can report the slice beside the official metric and reproduce the supplement (Section~\ref{sec:artifact}).
\end{itemize}

Our claim is narrow. Without adjudicating which track is ``correct'' --- the generator's own constraints can force these conflicts (Section~\ref{sec:benchmark}) --- one auditable request family exposes a measurable gap between policy-selected labels and visible user requests, and training can close it at no cost to the official metric.

\section{Policy-Selected Labels and the Audited System}
\label{sec:benchmark}

TalkPlayData 2 generates multi-turn music conversations agentically~\cite{choi2025talkplaydata2}: a Listener LLM, conditioned on a session goal and an LLM-inferred listening profile, converses with a Recsys LLM that must select every recommendation from a per-session in-context pool of 16--32 tracks drawn from one real user's listening session, with repeats forbidden across turns and the pool disjoint by construction from the profiling tracks that seed the listener's persona. The logged conversation is the dataset; the pool track the generator selected after each user turn is the official label.

The label is \emph{policy-selected}: it records the pipeline's continuation, which may serve the session goal, the flow of the dialogue, or the inferred profile that conditions both agents, rather than the user's latest visible request. The constrained pool sharpens this divergence: when the listener names a track outside the remaining pool, the logged label is structurally forced to be a substitute. The resulting conflicts need not be annotation errors of the kind cataloged in test-set audits~\cite{northcutt2021pervasive} --- the generator was serving its own session goal under its own constraints --- but they are supervision that a deployed system should not imitate.

The rest of the generation context is internal by design: the challenge asks systems to rank the full catalog, not to recover the pool, so the recommendation pool and profiling tracks stay inside the pipeline, and the released user profile omits attributes the profiling step produced, such as the user's top artist and genre. Everything in this paper is computed from visible dialogue and catalog metadata alone.

\textbf{Our challenge system.} All experiments in this paper run on our challenge system, a standard cascade. Retrieval is multi-arm. Three lexical BM25 arms search the dialogue text: one over the raw chat history, one pairing the current user message with the catalog metadata text of the last recommended track, and one pairing all user turns with the metadata text of every track recommended so far. Three dense arms search the organizer-provided embeddings of track metadata, lyrics, and tag attributes with a query vector encoded from the user turns. Four pseudo-relevance-feedback arms average the embeddings of BM25's top hits into a centroid and retrieve its nearest neighbors, in the three text spaces and in album-artwork space, which has no text-query arm of its own. One arm is a bi-encoder fine-tuned on dialogue-to-track pairs from the organizer training split. The union of each arm's top-200, deduplicated, gives about 1{,}550 candidates per turn. An XGBoost LambdaRank reranker~\cite{chen2016xgboost} scores each candidate with 49 features in six families: where the arms placed it, how it matches the query text, how it relates to the session's earlier tracks, how common its artist and album are within the turn's candidate pool, its catalog popularity, and the turn position. We call this model the base reranker. The final leaderboard submission added session co-occurrence features and a score-diffusion step to this cascade. An LLM writes the conversational response; its quality is scored separately by the challenge~\cite{doh2026llmjudge} and is out of scope here. All label interventions act on the reranker's training targets (Section~\ref{sec:training}).

\section{Related Work}
\label{sec:related}

\textbf{Simulated users and synthetic CRS evaluation.} Static single-label logs understate and distort conversational recommender system (CRS) quality~\cite{wang2023rethinking,bernard2025limitations}; LLM user simulators can leak the target item into the dialogue~\cite{zhu2024reliable}; much of reported CRS accuracy comes from re-recommending items already mentioned~\cite{he2023zeroshot}; and ReDial studies do not agree on which item counts as the gold label~\cite{kostric2026redial}. Our finding is complementary and, to our knowledge, new: in a fully synthetic benchmark the label can \emph{contradict} the visible request at a measurable rate, the opposite failure mode from leakage. Generation-time guards target exactly this failure --- round-trip consistency filtering~\cite{dai2023promptagator}, engineered label-dialogue consistency~\cite{liang2024llmredial,ryu2025icer} --- but they run inside the generator, on generator-trusted signals; we measure, post hoc and from participant-visible evidence alone, how often a shipped benchmark's labels still conflict with the visible request.

\textbf{Label and benchmark validity.} Label errors pervade canonical test sets~\cite{northcutt2021pervasive}, recommender evaluation has a history of illusory progress~\cite{dacrema2019progress}, and when an LLM produces the labels their validity conditions must be stated, not assumed~\cite{faggioli2023perspectives}. Dialogue benchmarking's standard remedy is re-annotation: MultiWOZ 2.1, 2.2, and 2.4 successively corrected state labels that misrecorded the dialogue~\cite{eric2020multiwoz21,zang2020multiwoz22,ye2022multiwoz24}. Our labels misrecord nothing --- they faithfully log the policy-selected continuation (Section~\ref{sec:benchmark}) --- so re-annotation is not the remedy; the intervention only supplements targets, never removes the official label (Section~\ref{sec:design}). Counterfactual learning likewise reads a logged label as the policy's choice rather than the user's intent~\cite{schnabel2016treatments,chaney2018confounding}, but corrects exposure bias in aggregate, with no per-turn request to check any label against; denoising methods adjust suspect training labels, but infer noise from loss dynamics or model agreement, not from anything the user visibly asked~\cite{wang2021denoising}. The nearest CRS neighbors treat the single label as \emph{incomplete}: Fashion-AlterEval widens a test set's relevance sets with new human judgments, at evaluation only~\cite{vlachou2025alterEval}; Xu et al.\ augment training labels with LLM-proposed relevant items, to counter false negatives~\cite{xu2025beyondsingle}. We audit the case neither covers: the visible request resolves against the catalog to a specific item that the official label \emph{contradicts}; the evidence is the catalog's resolution of the request, not a model's relevance judgment, and the fix must beat a matched control and preserve the official metric.

\textbf{Behavioral testing.} CheckList established behavior-level test suites that reveal failures invisible to aggregate accuracy~\cite{ribeiro2020checklist}, and RecList ported the idea to recommenders~\cite{chia2022reclist}. Those suites test the system; ours also tests the benchmark: we check the official labels against the visible request, and retrain on what we find.

\textbf{Exact-item intent and instruction following.} Honoring explicit user statements is increasingly framed as instruction following for recommenders, where both LLMs and retrieval models measurably fail~\cite{zhang2023instructrec,zhao2025prefeval,weller2024followir}. The benchmark's own authors' intent catalog treats a request for a named track the same as a genre or mood filter --- there is no exact-item category~\cite{doh2024musicdialogue}. Our audit makes that class first-class for one benchmark.

\section{Audit Design}
\label{sec:design}

Every step of the audit uses only what a deployed system can see: detection and every deployment-time action use visible dialogue and catalog metadata --- never the pipeline's logged reasoning traces --- and the official label is never an input to detection. The intervention may only \emph{supplement} or compare against the official label, never remove it: we add what the user visibly asked for without adjudicating which item is ``correct''.

Acting on a request family requires clearing a three-part evidence bar. The request must resolve from catalog metadata to a concrete target set, with explicit abstention on ambiguity and a detector that passes a strict precision audit. The trained model must beat a matched control that sees the same request \emph{feature} but not the request-satisfying targets, so the gain cannot be mere feature exposure. And the intervention must improve the request-satisfying readout while holding the official metric inside a predeclared preservation margin. Section~\ref{sec:families} reports the five visible families this bar declined.

\section{The Exact/Version Request Family}
\label{sec:family}

\textbf{The family.} The admitted family covers turns whose latest user message visibly requests a specific song title, or a version/duplicate equivalent of one: the same recording under an alternate catalog entry, such as a remaster or re-release.

\textbf{Detector.} Detection is deterministic --- regular expressions plus catalog lookup, no learned components --- so the audit is exactly reproducible and free to rerun. From the latest user turn the detector extracts quoted titles and imperative request spans (``play \ldots'', ``find \ldots'', ``can you play \ldots''), normalizing punctuation and version suffixes for matching only; a resolver then maps each extracted title to catalog tracks through title and artist fields, using an artist hint when the dialogue provides one. The detector abstains when no catalog item matches, when unrelated items tie, or when the language is broad preference rather than a directive. A learned or LLM detector would likely widen coverage; we leave that to future work.

\textbf{Prevalence and conflicts.} On the organizer development set (1{,}000 sessions; 8{,}000 evaluated turns) the detector finds 82 turns with a visible exact directive (a message that actively asks for a specific named track): 1.0\% of evaluated turns, a family rare in this benchmark though dominant in real usage (Section~\ref{sec:intro}). In 41 of the 82 the official label is a different track than the request resolves to, and conflicts concentrate deeper in the conversation --- from a sixth of turn-1 directives to over two-thirds by turns 5--8 --- the empirical signature of pool depletion under the no-repeat constraint (Section~\ref{sec:benchmark}).

\textbf{Detector validation.} We audited the detector against an independent LLM annotation pass (GPT-5.5), blind to the detector's output, on a stratified 120-turn sample from the organizer training split. Of 40 detector positives, 37 were explicit track requests, and in each the annotation named the same title the detector had extracted. The errors are mentions mistaken for requests, as in ``find more tracks that feature specific names, like `Chris R.' or `Sarah Sitting'\,'', where the quoted title follows \emph{find} but names an example, not a request. The detector also misses requests; the annotation pass found 3 in 80 sampled turns it had not flagged. The two error directions are not symmetric, and the audit is built for precision: a miss only means the 82 directives above undercount the family, while a false fire could inflate the conflict rate.

\section{Training with Request-Satisfying Targets}
\label{sec:training}

\textbf{Intervention.} Where the detector resolves an exact/version request to a target the official label does not already satisfy, the turn \emph{retains} its official label and adds the request-satisfying target set, with the whole turn's group weight reduced to 0.1 --- the two together are the supplement; all other turns are untouched and train at full weight, so the official supervision still dominates. At full training scale the action is small: 1{,}873 of 121{,}592 training turns are affected, contributing 1{,}878 additional positive rows (13 turns contribute none because the resolved track is absent from the turn's candidate pool; 17 contribute two or three because the request resolves to multiple catalog entries). We call the model retrained on this augmented supervision the specialist. A deployment-time alternative exists within these constraints: when the detector fires, promote the resolved track to rank~1. We study the training-target action instead, for two reasons: the override's slice score is near 1 by construction (the readout credits the same resolved targets the override promotes), so it says nothing about whether request-satisfying supervision conflicts with the official objective; and the override commits every detector error as a top-1 recommendation, while a trained reranker can weigh the request signal against other evidence.

\textbf{Matched control.} The comparison model is identical in every respect (candidate pools, deployment gate, training schedule, evaluation procedure, and all features: the base reranker's 49 plus an added binary exact-request feature) but trains on official labels only, at full weight. This isolates the supplement's contribution: if the specialist wins only because it knows a turn contains an exact request, the control would win equally. Both models are deployed behind the same inference-time dialogue gate: on turns where the visible-request detector fires, each retrained model's scores replace the base reranker's, and all other turns keep base scores. Control and specialist therefore share identical scores on non-gated turns, so every whole-split delta in the results originates from gated turns.

\textbf{Protocol freeze.} We built the detector, features, and target changes on internal splits of the training data, tuned them in a reduced-scale pilot read out on the development set, and then froze everything --- one predeclared configuration and readout, recorded in a manifest that ships with the artifact --- before both models were retrained on the full organizer training split and the evaluation below ran once.

\textbf{The evaluation slice.} The slice collects every directive where honoring the request and honoring the label diverge: the detector names the catalog tracks that satisfy the request, and the official label does not fully satisfy it. Directives whose label already satisfies the request contribute nothing. That gives 43 cases (42 exact-track, 1 version/duplicate), two more than the 41 conflicts of Section~\ref{sec:family}: an either/or directive whose label satisfies only one of its two named titles, and a version request admitted as the same recording under another catalog entry.

\section{Results}
\label{sec:results}

\begin{table}[t]
\caption{The 82 audited directives, by label status and the rank of the requested track under the matched control and the specialist, both deployed behind the gate. ``Absent'' means outside the model's top-20; for every track the specialist recovers, it was in the candidate pool throughout --- the loss is a ranking failure, not a retrieval one.}
\label{tab:ranks}
\small
\setlength{\tabcolsep}{2.5pt}
\centering
\begin{tabular}{@{}lr@{\,$\rightarrow$\,}l@{\hspace{2.2em}}r@{\,$\rightarrow$\,}l@{}}
\toprule
Requested track & \multicolumn{2}{c}{Label agrees (41)} & \multicolumn{2}{c}{Label conflicts (41)} \\
 & \multicolumn{2}{c}{control $\rightarrow$ specialist} & \multicolumn{2}{c}{control $\rightarrow$ specialist} \\
\midrule
Rank 1 & 33 & 41 & 13 & 25 \\
Ranks 2--20 & 8 & 0 & 10 & 15 \\
Absent & 0 & 0 & 18 & 1 \\
\bottomrule
\end{tabular}
\end{table}

Table~\ref{tab:ranks} follows all 82 audited directives --- not only the 43-case conflict slice --- under the two models: the specialist ranks the requested track first on 66 directives, the control on 46; turn by turn, the specialist ranks the requested track higher on 35 directives and lower on none. On the label-agreeing half the request is honored in full (41 of 41 at rank~1); on the conflict half the recovered tracks come at no measured cost to the official metric on the split as a whole (row one of Table~\ref{tab:main}). The one track still outside the specialist's top-20 belongs to a directive that is itself a detector false fire: the quoted phrase names a listening vibe, not a track.

Table~\ref{tab:main} reports three readouts: official nDCG@20 over the whole split; request-satisfying nDCG@20 over the whole split, which on audited turns accepts either the official label or the resolved target set as relevant; and the same request-satisfying readout averaged over only the 43-case conflict slice. The predeclared preservation margin is a whole-split official delta above $-0.001$.

\begin{table}[t]
\caption{Frozen full-train readout on the organizer development set: specialist vs.\ matched official-label control. The exact/version conflict slice is defined in Section~\ref{sec:training}.}
\label{tab:main}
\small
\setlength{\tabcolsep}{2.5pt}
\begin{tabular*}{\columnwidth}{@{\extracolsep{\fill}}lrrr@{}}
\toprule
Readout & Control & Specialist & $\Delta$\% \\
\midrule
Official nDCG@20 (whole split) & 0.1908 & 0.1914 & +0.31\% \\
Request-satisfying nDCG@20 (whole split) & 0.1922 & 0.1943 & +1.09\% \\
Request-satisfying nDCG@20 (conflict slice) & 0.5231 & 0.8018 & +53.3\% \\
\bottomrule
\end{tabular*}
\end{table}

The official metric is preserved (row 1): the whole-split delta is $+0.0006$ ($+0.31\%$), and a paired bootstrap over the 1{,}000 development sessions (10{,}000 resamples; all intervals are 95\% percentile CIs) puts the CI at $[+0.0002,+0.0012]$ --- inside the margin and excluding loss. Request satisfaction improves on the same ranked lists (row 2): $+1.09\%$ whole-split (CI $[+0.0011,+0.0032]$). The improvement concentrates exactly where label and request diverge (row 3): slice nDCG@20 rises from 0.523 to 0.802, a 53.3\% relative gain. A paired bootstrap over the 29 sessions that contain the 42 exact-track cases (the one version/duplicate case cannot be resampled) gives a session-level nDCG@20 delta of $+0.277$, CI $[+0.163, +0.400]$, positive in all 10{,}000 resamples. The control's 0.523 does not mean it half-honors requests --- the readout also credits the official label, which the control ranks well; the specialist's gain is on the request side.

The small whole-split deltas and the large slice delta are not in tension: the intervention touches 1.5\% of training turns and the conflict slice is 43 cases, so global movement is necessarily modest, while behavior changes sharply exactly where the request family applies. This is why the slice readout is needed: exact-song requests dominate real music interaction (Section~\ref{sec:intro}) yet occupy so few benchmark turns that the whole-split average barely registers how they are handled. The matched control rules out the explanation that the specialist merely learned that a turn contains an exact request.

\section{Other Request Families}
\label{sec:families}

Exact/version is the only family we act on; five other visible families each fail the evidence bar of Section~\ref{sec:design}. Hard artist constraints (``anything by this artist'') fail on precision: on training-split turns our artist detector reached 74\% action precision on strict review, short of the bar. Rejection and switch-away turns (``something else, please'') fail on preservation: suppressing the rejected track produces a statistically confirmed official loss ($-0.0009$, 95\% CI $[-0.0018,-0.0002]$) with no request-satisfying gain, and no audited switch turn has enough admissible alternatives for a clean top-20 (0/28). Album and year constraints fail target resolution: album strings match fuzzily, and release-date fields conflate original and reissue dates. Broad semantic requests --- descriptions by mood or attribute (``upbeat'') rather than by name, the most frequent family --- fail certifiability: none of 125 screened requests yielded a target set the catalog could verify. Declining is the audit working as designed: a family enters training only when its evidence clears the full bar.

\section{Reproducibility and Artifact}
\label{sec:artifact}

We release, under the code repository linked below: (i) ID-keyed audit annotations --- session, turn, request family, resolved target track ID, official label ID, and verdict --- for all audited development turns; (ii) the exact/version detector and its stratified precision audit; (iii) the target-remap script that reconstructs the request-satisfying evaluation slice from a user's own copy of the challenge data; and (iv) the training scripts for the specialist and matched-control rerankers. No dialogue text is redistributed, following the annotations-over-restricted-data pattern of re-annotated vision benchmarks~\cite{northcutt2021pervasive}. The evidence package adds the frozen protocol, training summaries, control readouts, bootstrap outputs, and the family-search ledger, with paths and hashes recorded in a manifest.

\begin{sloppypar}
Challenge metadata: team \teamname; official final score (Blind-B leaderboard): \scorenumbers, rank 12 of 44~\cite{doh2026recsyschallenge}. The training supplement of Section~\ref{sec:training} was not part of the leaderboard submission. Code: \codeurl
\end{sloppypar}

\section{Conclusion}
\label{sec:conclusion}

A recommender that ignores an exact-song request fails in the way users forgive least. Naming the desired item is the dominant intent in real music interaction, and deployed systems avoid substituting an alternative for a named item on the premise that it costs satisfaction --- yet in the RecSys Challenge 2026 TalkPlay benchmark, the official label points away from the visibly requested track in half of the 82 audited development turns, so a system trained to the leaderboard learns exactly this behavior. The failure hides from the official readout: exact directives are 1.0\% of evaluated turns, so the whole-split average barely moves however they are handled. It is also cheap to remove: supplementing 1.5\% of training turns with request-satisfying targets raises the conflict slice from 0.523 to 0.802 nDCG@20, beats a matched control, and preserves the official metric.

The scope is deliberate: the detector abstains when no catalog item resolves, so 41/82 describes the audited turns, not the benchmark at large. The detector is TalkPlay-specific; the evidence standard is not, and two extensions follow directly --- porting the audit to other synthetic CRS benchmarks, and running it inside the generation loop, where a label that contradicts the visible request could be caught before it is logged. We release the annotations and tooling so this slice can be reported beside the official metric.

\clearpage
\bibliographystyle{ACM-Reference-Format}
\bibliography{request_proxy_boundary_challenge_refs_v11}

\end{document}